\documentclass{skaox2006}
\usepackage{graphicx}
\begin{document}
   \title{Magnetism in Nearby Galaxies, Prospects with the SKA,
   and Synergies with the E-ELT}

   \author{R. Beck\inst{1}}

   \institute{Max-Planck-Institut f\"ur Radioastronomie, Auf dem H\"ugel
69, 53121 Bonn, Germany}

   \abstract{Radio synchrotron emission, its polarization and its Faraday
rotation are powerful tools to study the strength and structure of
interstellar magnetic fields. In the Milky Way, the total field
strength is about $6~\mu$G near the Sun and 50--100~$\mu$G near the
Galactic Center. Faraday rotation of the polarized emission from
pulsars and background sources indicate that the regular field
follows the spiral arms and has one reversal inside the solar
radius, but the overall field structure in our Galaxy is still
unclear. In nearby galaxies, turbulent fields are strongest in
spiral arms and bars (20--30~$\mu$G) and in central starburst
regions (50--100~$\mu$G). Ordered fields with spiral structure exist
in grand-design, barred and flocculent galaxies. The strongest
ordered fields (10--15~$\mu$G) are found in interarm regions.
Faraday rotation of the diffuse polarized radio emission from the
disks of spiral galaxies sometimes reveals large-scale patterns,
which are signatures of regular fields generated by a mean-field
dynamo. Ordered magnetic fields are also observed in radio halos
around edge-on galaxies, out to large distances from the plane, with
X-shaped patterns. -- The SKA and its precursor telescopes will open
a new era in the observation of cosmic magnetic fields and help to
understand their origin. The SKA will map interstellar fields in
nearby galaxies in unprecedented detail. All-sky surveys of Faraday
rotation measures (RM) towards a dense grid of polarized background
sources with the ASKAP (POSSUM), MeerKAT and the SKA are dedicated
to measure fields in intervening galaxies and will be used to model
the overall structure and strength of the magnetic fields in the
Milky Way and beyond. Examples for joint polarimetric observations
between the SKA and the E-ELT are given. }
   \maketitle
%
%

\section{Origin of cosmic magnetism}

Magnetic fields are a major agent in the ISM and also control the
density and distribution of cosmic rays. Cosmic rays accelerated in
supernova remnants can provide the pressure to drive galactic
outflows and buoyant loops of magnetic fields via the Parker
instability. Outflows from starburst galaxies in the early Universe
may have magnetized the intergalactic medium.

In spite of our increasing knowledge of cosmic magnetic fields, many
important questions, especially their origin and evolution, their
strength in intergalactic space, their first occurrence in young
galaxies and their dynamical importance for galaxy evolution remain
unanswered. The detection of ultrahigh-energy cosmic rays (UHECRs)
with the AUGER observatory calls for a detailed knowledge of the
magnetic field in the Milky Way to properly model the particle
propagation.

The most promising mechanism to sustain magnetic fields in the
interstellar medium of galaxies is the dynamo. In young galaxies a
small-scale dynamo (Brandenburg \& Subramanian \cite{brand05})
possibly amplified seed fields from the protogalactic phase to the
energy density level of turbulence within less than $10^9$~yr. To
explain the generation of large-scale fields in galaxies, the
mean-field dynamo has been developed (Beck et al. \cite{beck96}). It
is based on turbulence, differential rotation and helical gas flows,
generated by supernova explosions (Gressel et al. \cite{gressel08})
or by cosmic-ray driven Parker loops (Hanasz et al.
\cite{hanasz09}). The mean-field dynamo in galaxy disks predicts
that within a few $10^9$~yr large-scale regular fields are excited
from the seed fields (Arshakian et al. \cite{arshakian09}), forming
patterns (``modes'') with different azimuthal symmetries in the disk
and vertical symmetries in the halo.

%
%

\section{Observation of cosmic magnetism}

\begin{table*}[width=\textwidth]
  \begin{center}
  \caption{\label{tab:obs}
  Comparison of polarization observations in the optical/near-infrared
  and in the radio continuum range}
  \begin{tabular}{lll}
  \hline
     & Optical / infrared & radio continuum \\
    \hline
     Detectors: & special analyzers & normal receivers \\
     {\bf A.} Continuum emission processes: & scattering, dust, synchrotron & synchrotron  \\
     Sources: & ISM, galactic nuclei, jets & ISM, SNRs, galactic
     nuclei, jets \\
     Degrees of linear polarization: & a few \% & up to 70\% \\
     Wavelength dependence: & weak & strong at long $\lambda$ (Faraday depolarization) \\
     Spectro-polarimetry: & N/A & under development \\
     {\bf B.} Line emission processes: & Zeeman, Hanle, AMR & Zeeman \\
     Sources: &  sun, stars, planets  & ISM \\
     Spectro-polarimetry: & established & N/A \\
   \hline
  \end{tabular}
  \end{center}
\end{table*}

Magnetic fields need illumination to be detectable. {\em Polarized
emission}\ at optical, infrared, submillimeter and radio wavelengths
holds the clue to magnetic fields in galaxies. Optical linear
polarization is a result of extinction by elongated dust grains in
the line of sight which are aligned in the interstellar magnetic
field (the {\em Davis-Greenstein effect}). The E--vector runs
parallel to the field. However, light can also be polarized by
scattering, a process unrelated to magnetic fields and hence a
contamination that is difficult to subtract from the diffuse
polarized emission from galaxies, e.g. in M~51 (Scarrott et al.
\cite{scarrott87}). Optical polarization data of about 5500 selected
stars in the Milky Way yielded the orientation of the magnetic field
near the Sun (Fosalba et al. \cite{fosalba02}). Together with
measurements of stellar distances, a 3-D analysis of the magnetic
field within about 5~kpc from the Sun is possible, but more data are
needed.

Linearly polarized emission from elongated dust grains at infrared
and submillimeter wavelengths is not affected by scattered light.
The B--vector is parallel to the magnetic field. The field structure
can be mapped in gas clouds of the Milky Way (e.g. Tang et al.
\cite{tang09}) and in galaxies, e.g. in the halo of M~82 (Greaves et
al. \cite{greaves00}).

In the optical and near-infrared lines, the Zeeman and Hanle effects
are the main polarization mechanisms, and spectro-polarimetric
observations (``probing the third dimension'', Oudmaijer \& Harries
\cite{oudmaijer08}) have enormously increased our knowledge about
magnetic fields in stars (Beryugina \cite{berdyugina09}). The {\em
Atomic Magnetic Realignment (AMR)}\ (Yan \& Lazarian \cite{yan08})
mechanism predicts a few \% polarization e.g. in the optical and UV
lines of O{\sc i}, S{\sc ii} and Ti{\sc ii}, but has not yet been
detected.

Most of what we know about interstellar magnetic fields comes
through the detection of radio waves. {\em Zeeman splitting}\ of
radio spectral lines directly measures the field strength in gas
clouds of the Milky Way (Heiles \& Troland \cite{heiles05}) and in
starburst galaxies (Robishaw et al. \cite{robishaw08}). The
intensity of {\em synchrotron emission}\ is a measure of the number
density of cosmic-ray electrons in the relevant energy range and of
the strength of the total magnetic field component in the sky plane.
The assumption of energy equipartition between these two components
allows us to calculate the total magnetic field strength from the
synchrotron intensity (Beck \& Krause \cite{beck+krause05}).

Polarized emission emerges from ordered fields. As polarization
``vectors'' are ambiguous by $180\degr$, they cannot distinguish
{\em regular (coherent) fields}, defined to have a constant
direction within the telescope beam, from {\em anisotropic fields},
which are generated from turbulent fields by compressing or shearing
gas flows and frequently reverse their direction along the other two
dimensions. Unpolarized synchrotron emission indicates {\em
turbulent (random) fields}\, which have random directions in 3-D and
have been amplified and tangled by turbulent gas flows.

The intrinsic degree of linear polarization of synchrotron emission
is about 75\%. The observed degree of polarization is smaller due to
the contribution of unpolarized thermal emission, which may dominate
in star-forming regions, by {\em Faraday depolarization}\ along the
line of sight and across the beam (Sokoloff et al.
\cite{sokoloff98}), and by geometrical depolarization due to
variations of the field orientation within the beam.

At radio wavelengths of a few centimeters and below, the orientation
of the observed B--vector is parallel to the field orientation, so
that the magnetic patterns of many galaxies can be mapped directly
(e.g. Beck \cite{beck05}). At longer wavelengths, the polarization
vector is rotated in a magnetized thermal plasma by {\em Faraday
rotation}. The rotation angle increases with the square of the
wavelength $\lambda^2$ and with the {\em Rotation Measure (RM)},
which is the integral of the plasma density and the strength of the
component of the field along the line of sight. As the rotation
angle is sensitive to the sign of the field direction, only regular
fields give rise to Faraday rotation, while anisotropic and random
fields do not. Measurements of the Faraday rotation from
multi-wavelength observations allow to determine the strength and
direction of the regular field component along the line of sight.
Dynamo modes of regular fields can be identified from the pattern of
polarization angles and RMs of the diffuse polarized emission of
galaxy disks (e.g. Fletcher et al. \cite{fletcher04}).

Distinct emitting regions located on the line of sight can generate
several RM components, so that the observed RM is no longer a linear
function of $\lambda^2$. In such cases, multi-channel
spectro-polarimetric radio data are needed that can be
Fourier-transformed into Faraday space, called {\em RM Synthesis}\
(Brentjens \& de Bruyn \cite{brentjens05}, Heald \cite{heald09}). If
the medium has a relatively simple structure, the 3-D structure of
the magnetized interstellar medium can be determined ({\em Faraday
tomography}).

A grid of RM measurements towards polarized background sources is a
powerful tool to study magnetic field patterns in galaxies (Stepanov
et al. \cite{stepanov08}). A large number of background sources is
required to recognize the field patterns, to separate the Galactic
foreground contribution and to account for intrinsic RMs of the
extragalactic sources.

Table~\ref{tab:obs} summarizes the main properties of the
observation methods.

%
%

\section{Nearby Universe: Projects with the Square Kilometre Array}

The SKA will allow us to investigate the interstellar medium (ISM)
in nearby galaxies with much higher resolution and/or much higher
sensitivity than with present-day telescopes. Several projects are
discussed (e.g. Carilli \& Rawlings \cite{carilli04}):
\begin{itemize}
\item HI line: Turbulence spectrum of the ISM in the Milky Way
down to scales of a few AU
\item HI line: ISM structure $\le100$~pc in galaxies up to
$\approx20$~Mpc distance
\item HI line: Low-density gas in outer parts of galaxies
\item HI line: High-velocity clouds around galaxies and in
the IGM
\item Zeeman effect (HI, OH and H$_2$O lines): Magnetic fields in
nearby galaxies
\item Thermal emission: Structure of ultracompact HII regions up
to $\approx1$~Mpc distance and super star clusters up to
$\approx20$~Mpc distance
\item Thermal emission: Stellar winds and stellar jets
\item Recombination lines: Structure of thermal gas in nearby
galaxies
\item Methanol lines: Parallaxes of masers in nearby galaxies
\item Synchrotron emission: Structure and variability of knots in
SNRs and of galactic nuclei
\item Pulsars in the Milky Way and nearby galaxies: Testing theories
of gravity and detection of gravitational waves
\item Polarized synchrotron emission and Faraday rotation: Origin
and evolution of cosmic magnetism.
\end{itemize}

The last two topics with underlying fundamental physical questions
were chosen as Key Science Projects. The pulsar project is described
elsewhere (Kramer, this volume). A summary of the SKA magnetism
project is given in Section~\ref{sec:ska}.

%
%

\section{Magnetic fields in the Milky Way}

Optical polarization data of stars at distances $>$1~kpc show that
the magnetic field of the Milky Way is predominantly oriented
parallel to the disk plane (Fosalba et al. \cite{fosalba02}). The
orientations of the polarization vectors from more nearby stars are
chaotic around $80^\circ$ Galactic longitude, which points to the
direction of the local field.

Surveys of the total synchrotron emission from the Milky Way yield
equipartition strengths of the total field of 6~$\mu$G, averaged
over about 1~kpc around the Sun, consistent with the HI Zeeman
splitting data of low-density gas clouds (Heiles \& Troland
\cite{heiles05}), and about 10~$\mu$G in the inner spiral arms
(Berkhuijsen, in Wielebinski \cite{wielebinski05}). Faraday and
dispersion measure data of pulsars give an average strength of the
local regular field of $1.4\pm0.2~\mu$G, while in the inner Norma
arm the average strength of the regular field is $4.4\pm0.9~\mu$G
(Han et al. \cite{han02}).

In the synchrotron filaments near the Galactic center, oriented
almost perpendicular to the plane, a break in the spectrum indicates
that the field strength is 50--100~$\mu$G (Crocker et al.
\cite{crocker10}).

\begin{figure}
\centering
\includegraphics[bb = 48 164 521 630,width=6cm,clip=]{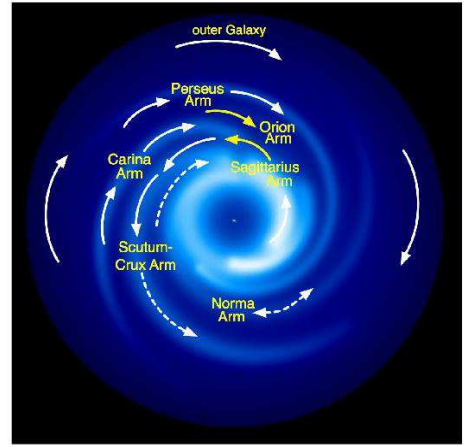}
\caption{Model of the magnetic field in the Milky Way, derived from
Faraday rotation measures of pulsars and extragalactic sources.
Generally accepted results are indicated by yellow vectors, while
white vectors refer to results which are probable but not fully
confirmed (from Brown, priv. comm.). } \label{fig:galaxy}
\end{figure}

The all-sky maps of polarized synchrotron emission at 1.4~GHz from
the Milky Way from DRAO and Villa Elisa and at 22.8~GHz from WMAP
and the new Effelsberg RM survey of polarized extragalactic sources
were used to model the regular Galactic field (Sun et al.
\cite{sun08}). One large-scale field reversal is required at about
1--2~kpc from the Sun towards the Milky Way's center, which is also
supported by the detailed study of RMs from extragalactic sources
near the Galactic plane (Van Eck et al. \cite{eck10},
Fig.~\ref{fig:galaxy}). The overall structure of the regular field
is not known yet - its structure cannot be described by a simple
dynamo-type pattern (Men et al. \cite{men08}). A larger sample of
pulsar and extragalactic RM data is needed.

The signs of RMs of extragalactic sources and of pulsars at Galactic
longitudes l=90$^\circ$--270$^\circ$ are the same above and below
the plane (Taylor et al. \cite{taylor09}): the local magnetic field
is part of a symmetric field structure. In contrast, the RM signs
towards the inner Galaxy (l=270$^\circ$--90$^\circ$)  are {\em
opposite}\ above and below the plane. This can be assigned to an
antisymmetric halo field (Sun et al. \cite{sun08}) or to deviations
of the local field.

Little is known about the vertical field component in the Milky Way.
According to RM data from extragalactic sources, the local regular
Galactic field has no significant vertical component towards the
northern Galactic pole and only a weak vertical component of
$B_z\simeq0.3~\mu$G towards the south (Mao et al. \cite{mao10}),
while mean-field dynamo models predict vertical fields towards both
poles.

%
%

\section{Magnetic fields in spiral galaxies}

The typical average equipartition strength of the total magnetic
field in spiral galaxies is about $10~\mu$G. Radio-faint galaxies
like M~31 and M~33, our Milky Way's neighbors, have weaker total
magnetic fields (5--7~$\mu$G), while gas-rich spiral galaxies with
high star-formation rates, like M~51, M~83 and NGC~6946, have total
field strengths of 20--30~$\mu$G in their spiral arms. The strongest
total fields of 50--100~$\mu$G are found in starburst galaxies, e.g.
in M~82 (Klein et al. \cite{klein88}).

\begin{figure}
\centering
\includegraphics[bb = 47 213 522 573,width=7cm,clip=]{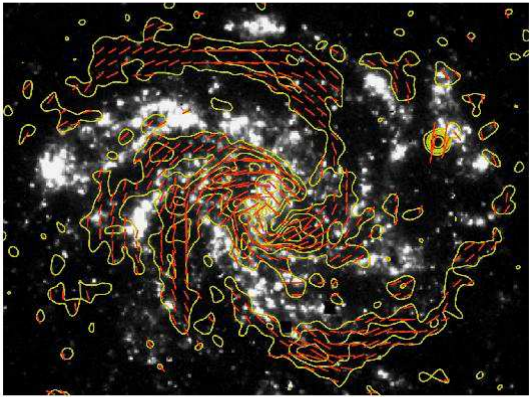}
\caption{Polarized radio emission (contours) and $B$--vectors of
NGC~6946, combined from observations at 6~cm wavelength with the VLA
and Effelsberg telescopes and smoothed to 15'' resolution (from Beck
\& Hoernes \cite{beck+hoernes96}). The background image shows the
H$\alpha$ emission (from Ferguson et al. \cite{ferguson98}).
Copyright: MPIfR Bonn. Graphics: \textit{Sterne und Weltraum}. }
\label{fig:n6946}
\end{figure}

The ordered
fields traced by the polarized synchrotron emission are generally
strongest (10--15~$\mu$G) in the regions {\em between}\ the optical
spiral arms and oriented parallel to the adjacent spiral arms, in
some galaxies forming {\em magnetic arms}\ (Fig.~\ref{fig:n6946}),
probably generated by a mean-field dynamo. In galaxies with strong
density waves some of the ordered field is concentrated at the inner
edge of the spiral arms (Fletcher et al. \cite{fletcher10}). The
ordered field forms spiral patterns in almost every galaxy (Beck
\cite{beck05}), even in ring galaxies (Chy{\.z}y \& Buta
\cite{chyzy08}) and in flocculent galaxies without massive spiral
arms (Soida et al. \cite{soida02}). Spiral fields
are also observed in the central regions of galaxies and in
circum-nuclear gas rings (Beck et al. \cite{beck+05}).

\begin{figure}
\centering
\includegraphics[bb= 26 27 235 358,width=5cm,angle=270]{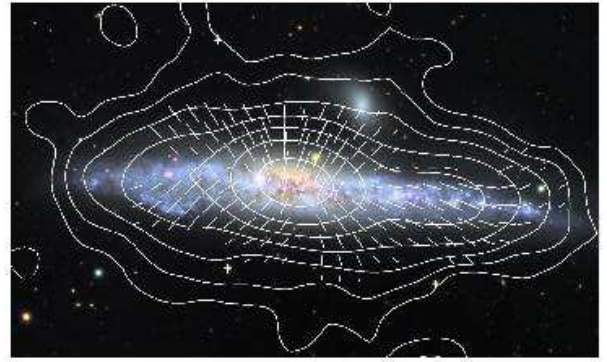}
\caption{Total radio emission (84'' resolution) and B--vectors of
the edge-on irregular galaxy NGC~4631, observed at 3.6~cm wavelength
with the Effelsberg 100-m telescope. The background optical image is
from the Misti Mountain Observatory. Copyright: MPIfR Bonn (from
Krause \cite{krause09}).} \label{fig:n4631}
\end{figure}

Spiral fields can be generated by compression at the inner edge of
spiral arms, by shear in interarm regions, or by dynamo action.
Large-scale patterns of Faraday rotation measures are signatures of
coherent dynamo fields and can be identified from polarized emission
of the galaxy disks (Fletcher et al. \cite{fletcher04}) or from RM
data of polarized background sources (Stepanov et al.
\cite{stepanov08}). The Andromeda galaxy M~31 and several other
galaxies host a dominating axisymmetric disk field, as predicted by
dynamo models. However, in many observed galaxy disks no clear
patterns of Faraday rotation were found. Either the field structure
cannot be resolved with present-day telescopes or the generation of
large-scale modes takes longer than the galaxy's lifetime (Arshakian
et al. \cite{arshakian09}).

Nearby galaxies seen edge-on generally show a disk-parallel field
near the disk plane. High-sensitivity observations of NGC~4631
(Fig.~\ref{fig:n4631}), NGC~253 (Heesen et al. \cite{heesen09}) and
other galaxies revealed ``X-shaped'' fields in the halo. The field
is probably transported from the disk into the halo by an outflow
emerging from the disk. Faraday RM data of NGC~253 indicate a
symmetric quadrupolar field, consistent with mean-field dynamo
models.

Polarized emission can also be detected from unresolved galaxies if
the inclination is larger than about $20\degr$ (Stil et al.
\cite{stil09}). This opens a new method to search for ordered fields
in distant galaxies.

%
%
\begin{table*}[width=\textwidth]
  \begin{center}
  \caption{\label{tab:syn} Synergies on magnetic field investigations
  in the Milky Way and nearby galaxies}
  \begin{tabular}{lll}
  \hline
     & SKA & E-ELT \\
    \hline
     Detailed field structure in the ISM & Sync pol & Optical or NIR pol \\
     Field structure in molecular clouds & Sync pol & NIR pol \\
     Field strengths in molecular clouds & Sync pol \& Zeeman & NIR pol \\
     Field ordering by gas flows & Sync pol \& HI velocities & Opt./NIR pol \& velocity fields \\
     Stellar jets & Sync pol & Emission lines \\
     Galactic Center \& AGNs & Sync pol & Opt./NIR pol \& emission lines \\
   \hline
  \end{tabular}
  \end{center}
\end{table*}

\section{Prospects with the SKA and its Precursors}
\label{sec:ska}

Future radio telescopes will widen the range of observable magnetic
phenomena. Low-frequency radio telescopes like the Low Frequency
Array (LOFAR) and the planned Murchison Widefield Array (MWA), Long
Wavelength Array (LWA) and the low-frequency part of the SKA will be
suitable instruments to search for extended synchrotron radiation at
the lowest possible levels in outer galaxy disks and the transition
to intergalactic space (Beck \cite{beck09}).

High-resolution, deep observations of the detailed field structure
in the ISM of galaxies at high frequencies, where Faraday
depolarization is small, require a major increase in sensitivity for
continuum observations that will be achieved by the EVLA and
especially by the SKA.

If polarized emission from galaxies is too weak to be detected, the
method of RM grids towards background sources (QSOs) can still be
applied. The {\em POSSUM}\ survey at 1.4~GHz planned with the
Australia SKA Pathfinder (ASKAP) telescope with 30~deg$^2$ field of
view (Gaensler et al. \cite{gaensler10}) will measure about 100 RM
values per square degree from polarized extragalactic sources within
10~h integration time. Measuring RM grids for a survey of nearby
galaxies has been proposed for MeerKAT, the SKA precursor in South
Africa. The SKA ``Magnetism'' Key Science Project plans to observe a
wide-field survey (at least $10^4$~deg$^2$) around 1~GHz with 1~h
integration per field, which will measure at least 1500 RMs per
deg$^2$ and a total RM number of at least $2\times10^7$ (Gaensler et
al. \cite{gaensler04}). More than $10^4$ RM values are expected in
the area of M~31 and will allow the detailed reconstruction of the
3-D field structure in this and many other nearby galaxies.
Large-scale field patterns can be recognized out to distances of
about 100~Mpc (Stepanov et al. \cite{stepanov08}).

The SKA pulsar survey will find about 20\,000 new pulsars which will
be mostly polarized and reveal RMs, perfectly suited to measure the
Milky Way's magnetic field with high precision (Noutsos
\cite{noutsos09}).

%
%

\section{Synergies with the E-ELT}

Polarization observations in radio continuum with the SKA and in the
optical or near-infrared with the E-ELT have a lot of synergy
potential (Table~\ref{tab:syn}). Polarization is a standard
observation mode for the SKA (Gaensler et al. \cite{gaensler04}),
but polarimetry modes with the E-ELT still need to be developed
(Strassmeier \& Ilyin \cite{strassmeier09}). The weak signals of
optical polarimetry need large collecting areas and hence can best
be done from ground, and photometric nights are not needed. {\em The
E-ELT needs polarimeters!}

%
%



\end{document}